**Астапеня Володимир Михайлович**
доцент кафедри інформаційної та кібернетичної безпеки
Київський університет імені Бориса Грінченка, Київ, Україна
OrcID: 0000-0003-0124-216X
*v.astapenia@kubg.edu.ua*

**Соколов Володимир Юрійович**
старший викладач кафедри інформаційної та кібернетичної безпеки
Київський університет імені Бориса Грінченка, Київ, Україна
OrcID: 0000-0002-9349-7946
*v.sokolov@kubg.edu.ua*

**Таджіні Махіяр**
молодший науковий співробітник кафедри інформаційної та кібернетичної безпеки
Київський університет імені Бориса Грінченка, Київ, Україна
OrcID: 0000-0001-8875-3362
*m.tajdini@kubg.edu.ua*

# РЕЗУЛЬТАТИ ТА ЗАСОБИ ОЦІНКИ ЕФЕКТИВНОСТІ СИСТЕМ ФОКУСУВАННЯ ДЛЯ ПІДВИЩЕННЯ ДОСТУПНОСТІ В БЕЗПРОВОДОВИХ МЕРЕЖАХ

**Анотація.** Широке поширення безпроводових технологій призводить до постійно зростання кількості користувачів і постійно функціонуючих пристроїв. Але зростання кількості безпроводових користувачів в обмеженому просторі і обмеженому частотному діапазоні призводить до зростання їх взаємного впливу, що в кінцевому підсумку негативно позначається на пропускній спроможності безпроводових каналів і навіть на працездатності системи в цілому. У статті наведено статистику і тенденції поширення безпроводових мереж систем стандарту IEEE 802.11, а також проаналізовано основні проблеми, що виникають в ході розширення їх використання. Обґрунтування і вибір шляхів подолання цих труднощів багато в чому залежить від об'єктивного контролю параметрів випромінювання точок доступу і абонентських коштів в конкретній обстановці. Наведено огляд штатних засобів контролю, передбачених розробниками обладнання, і запропоновані авторські варіанти експериментальних вимірювальних комплексів, що дозволяють контролювати сигнальні та інформаційні параметри систем Wi-Fi. Представлені отримані з використанням зазначених коштів експериментальні результати оцінки підвищення доступності та пропускної здатності на основі застосування прискорюючої металопластинчастої лінзи як додаткового автономного елементу для фокусування поля в тому числі і для систем MIMO, впливу прискорюючої металопластинчастої лінзи на просторовий розподіл поля, на спектральну структуру сигналу. Крім того, досліджувалися поляризаційні ефекти. Обговорюються можливі шляхи подальшого підвищення доступності, цілісності інформації та енергетичної ефективності систем безпроводового доступу. Автори пропонують більш прості і менш витратні варіанти підвищення спрямованості випромінювання на основі прискорюючої металопластинчастої лінзи, що випробувано експериментально, а також використання зонування простору на шляху ЕОМ.

**Ключові слова:** безпроводова мережа; металопластинчаста лінза; прискорююча лінза; лінзова антена; точка доступу; поляризація; MIMO.

## 1. ВСТУП

Незаперечною перевагою систем безпроводового доступу до інформаційних ресурсів, які набувають все більшого поширення, є зручності, що надаються користувачам. Зростає ступінь покриття території, збільшується швидкість передачі інформації, пропускна здатність системи, число точок доступу, щільність їх розміщення, особливо в містах, місцях розташування організацій і установ, в зонах відпочинку і т. п., що ілюструє статистика розвитку такої форми надання інформаційних послуг. Широке поширення безпроводових технологій призводить до постійно зростання кількості користувачів і постійно функціонуючих пристроїв. Але зростання кількості безпроводових користувачів в обмеженому просторі і обмеженому частотному діапазоні призводить до зростання їх взаємного впливу, що в кінцевому підсумку негативно позначається на пропускній спроможності безпроводових каналів і навіть на працездатності системи в цілому.

Окремим показником є період подвоєння кількості безпроводових мереж (тільки для стандарту IEEE 802.11) [1]. З графіка, показаного на рис. 1, видно, що середній час поновлення становить менше двох років і добре узгоджується з законом Мура. На графіку видно, що криза 2008 привів до затримки росту приблизно не три роки, яка потім компенсувалася прискореним зростанням кількості точок доступу. Якщо усереднити швидкість росту кількості (на графіку показано пунктирною лінією), то видно, що процес поступово сповільнюється, але все ж носить наростаючий характер.

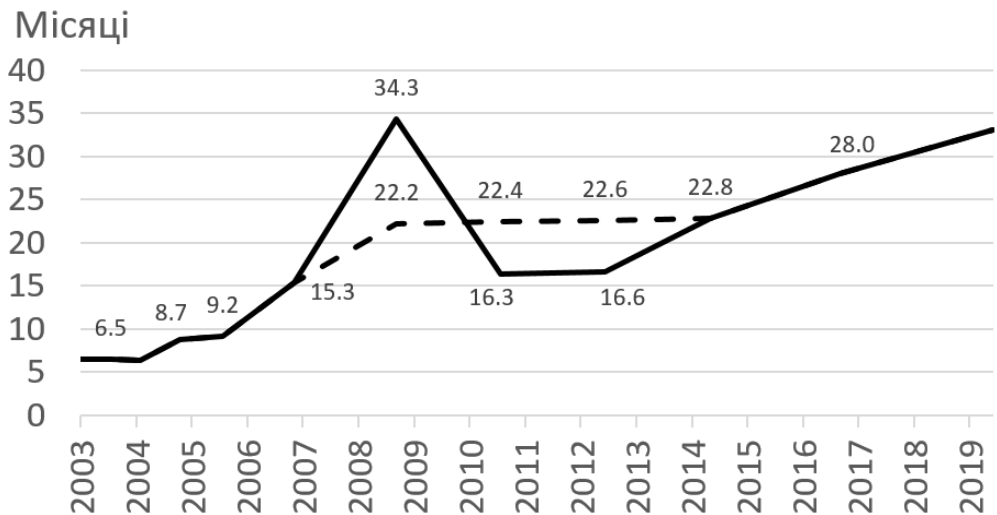

*Рис. 1. Терміни подвоєння кількості безпроводових точок доступу*

Графік дозволяє спрогнозувати наступне подвоєння кількості безпроводових мереж на початок 2020 року.

## 2. ОГЛЯД АПАРАТНОГО ЗАБЕЗПЕЧЕННЯ ДЛЯ АНАЛІЗУ СПЕКТРУ

Разом з тим, бурхливе зростання популярності і кількості безпроводових мереж (особливо в умовах щільної міської забудови) призводить до взаємного впливу мереж друг на друга. Крім того, збільшується кількість мобільних пристроїв і вбудованих систем, які можуть бути використані в якості безпроводових точок доступу (наприклад, модулі сімейства Espressif ESP8266, ESP32, Onion Omega 2, Realtek RTL8710, RDA5981 тощо). Це неминуче породжує свої проблеми і загрози, в тому числі і для безпеки інформації (порушення цілісності, проблеми з доступністю і т. ін.). До них можна віднести як банально ненавмисну перевантаження системи доступу, так і більш специфічні і навіть витончені погрози: несанкціонованому доступу до мережі; DoS-атаки; підробка службових пакетів; примусова деавтентифікація користувачів; сплески рівнів ненавмисних перешкод (у всьому використовуваному діапазоні або окремих частотних каналах); загроза постановки навмисних активних перешкод в умови конкурентної боротьби або терористичних дій. При цьому всі існуючі системи безпроводового доступу (стільниковий зв'язок, Wi-Fi, розумний будинок і інші) мають істотний недолік, — вони мають дуже низьку енергетичну ефективність, так як лише незначна частина випромінювання енергії сприймається приймальною антеною.

Певною мірою для подолання зазначених проблем можуть використовуватися такі методи.

1. Підвищення ефективного використання частотного ресурсу; для вирішення цього завдання:

– адміністративне централізоване планування безпроводової інфраструктури, законодавчі обмеження на рівень сигналу або потужність випромінювачів та ін.;

– регулярний моніторинг і адаптація систем вручну;

– застосування адаптивних систем підстроювання на рівні протоколів (IEEE 802.11f, IEEE 802.11k), моніторингу і автопідлаштування на рівні приймального тракту у деяких виробників (Atheros Spectral Scan mode), використання додаткових пристроїв для збору інформації про стан безпроводової системи.

2. Впровадження методів просторового (кутового) поділу абонентів на основі антен з вузькими і керованими характеристиками спрямованості аж до фазованих антенних решіток.

3. Видозміну і дообладнання середовища функціонування безпроводової системи доступу без істотної зміни конструкції обладнання точок доступу і абонентських засобів.

У зазначених умовах помітну роль в забезпеченні живучості інформаційної мережі можуть зіграти засоби моніторингу та контролю стану системи безпроводового доступу, що забезпечують об'єктивну оцінку параметрів системи і дозволяють виробити адекватні заходи щодо усунення причин зниження доступності та цілісності інформації як на етапі її розгортання і налагодження, так і для адаптації в реальному часі.

Подібні засоби представлені в лінійках продуктів для безпроводового доступу у Cisco, Ubiquiti, HP та ін. Багато виробників безпроводового обладнання вбудовують алгоритми автоматичного вибору каналів, але спектр сканується лише в області розташування точки доступу. При цьому не враховуються особливості розташування клієнтів. Точка доступу починає працювати на самому вільному каналі в місці її розташування, що в певній мірі покращує роботу всієї мережі, але не робить її оптимальною (так як неможливо врахувати всі параметри в ad hoc мережі: поляризацію, висоту розташування, екранування і перевідбиття, а також переміщення користувача). Тому крім інформації від точки доступу слід враховувати частотну обстановку і рівень сигналу в місці розташування користувачів (всіх або вибіркових) і, виходячи з отриманих даних, вибирати оптимальний частотний канал. Для збору інформації можна використовувати вже існуючі безпроводовий карти, але область їх видимості часто обмежена лише мережами стандарту IEEE 802.11 (а деякі карти навіть не бачать «прихованих» мереж); не враховується вплив інших безпроводових технологій, які працюють в тому ж частотному діапазоні, наприклад, за стандартами

IEEE 802.15.1 (Bluethooth), IEEE 802.15.4 (ZigBee, WirelessHART, MiWi, ISA100.11) та інших нестандартних пристроїв, а також побутових і промислових перешкод.

Однак ці кошти володіють не всі властивості і функціями, необхідними для аналізу показників тих удосконалень, які автори пропонують для підвищення доступності та цілісності інформації в системах безпроводового доступу стандарту IEEE 802.11.

Виникла необхідність використовувати додаткові незалежні пристрої контролю рівня сигналу, спектра випромінювання у всьому виділеному діапазоні і (бажано) пропускної здатності системи.

До цих пристроїв контролю пред'явлені такі вимоги:

1. Можливість розміщення в ключових точках інфраструктури і в місцях розташування користувачів.

2. Діаграма спрямованості антени повинна наближатися до сферичної (антена неспрямована).

3. Вимірювання рівня сигналу.

4. Аналіз спектру випромінювання у всьому виділеному діапазоні, а також спектра сигналу, використовуваного в експерименті.

5. Оцінка інформаційних характеристик каналу.

6. Для передачі результатів повинні використовуватися безпроводові або провідні канали зв'язку, які виходять за межі об'єкту сканування частотного діапазону.

7. Мінімальне енергоспоживання (для можливості автономного живлення).

Тому для забезпечення проведених (та перспективних) експериментальних робіт, авторами був розроблений, виготовлений і застосований вимірювальний комплекс контролю спектра і рівня випромінювання, а також необхідні для його функціонування програмні засоби аналізу як характеристик сигналу, так і інформаційних потоків.

Як аналізаторів спектра вибиралися пристрої з представлених на ринку (рис. 2). З усіх аналізаторів найбільш зручним виявився Pololu Wixel, так як він має добре продуману архітектуру, доступ до зміни прошивки і SDK (на мові програмування C) з повною документацією. Доступна ціна пристрою дозволяє одночасно запускати декілька датчиків, а даних з них збирати одночасно і паралельно аналізувати. Також було написано програмне забезпечення MDRV (на мові програмування Java) для одночасної роботи з декількома аналізаторами спектра.

## 3. ЗАСТОСУВАННЯ РОЗРОБЛЕНИХ ЗАСОБІВ КОНТРОЛЮ

В ході досліджень можливостей підвищення доступності інформації (і збільшення дальності доступу) в системах Wi-Fi на основі використання прискорюючої металопластинчастої лінзи (ПМЛ) на різних етапах застосовувалося різне апаратне забезпечення як для організації безпроводових точок доступу, так і для аналізаторів спектра [2]. У перших експериментах [3–10] був використаний комплект у складі:

– точки безпроводового доступу (передавач) Asus N16 (на базі мікроконтролера Broadcom 4718A, 533 МГц з прошивкою DD-WRT v24-sp2 mega і ненаправленої антеною);

– зовнішній безпроводової адаптер (приймач) на базі мікроконтролера Realtek RTL8187L з ненаправленою антеною. Аналіз спектру сигналу проводився за допомогою аналізатора спектра Ubiquiti AirView2 (див. рис. 2д).

Це дозволило оцінити ступінь підвищення пропускної спроможності з ПМЛ в порівнянні з випадком роботи промислового роутера без ПМЛ (за інших рівних умов).

В експериментах по оцінці можливості застосування ПМЛ в системах MIMO [8–10] система контролю була вдосконалена за рахунок використання промислового роутера зі схемою MIMO 2×3:2 з використанням двох (з трьох доступних) антен, які були вийняті з корпуса і додатково рознесені. Вимірювальний стенд включав в себе:

– точку безпроводового доступу (передавач) Asus N16 (на базі мікроконтролера Broadcom 4718A, 533 МГц з прошивкою DD-WRT K2.6 Big Generic rev. 14896);

– мережеву карту (приймач) Atheros AR9287 (MIMO 2×2:2);

– в ролі датчиків поля використані програмовані модулі Popolu Wixel (на базі мікроконтролера TI CC2511F32), показані на рис. 2ж.

Інформація оброблялася за допомогою спеціального програмного забезпечення MDRV, яке збирає дані з датчиків і в режимі реального часу відображає результат у зведеній формі на загальній діаграмі реального часу [8–10].

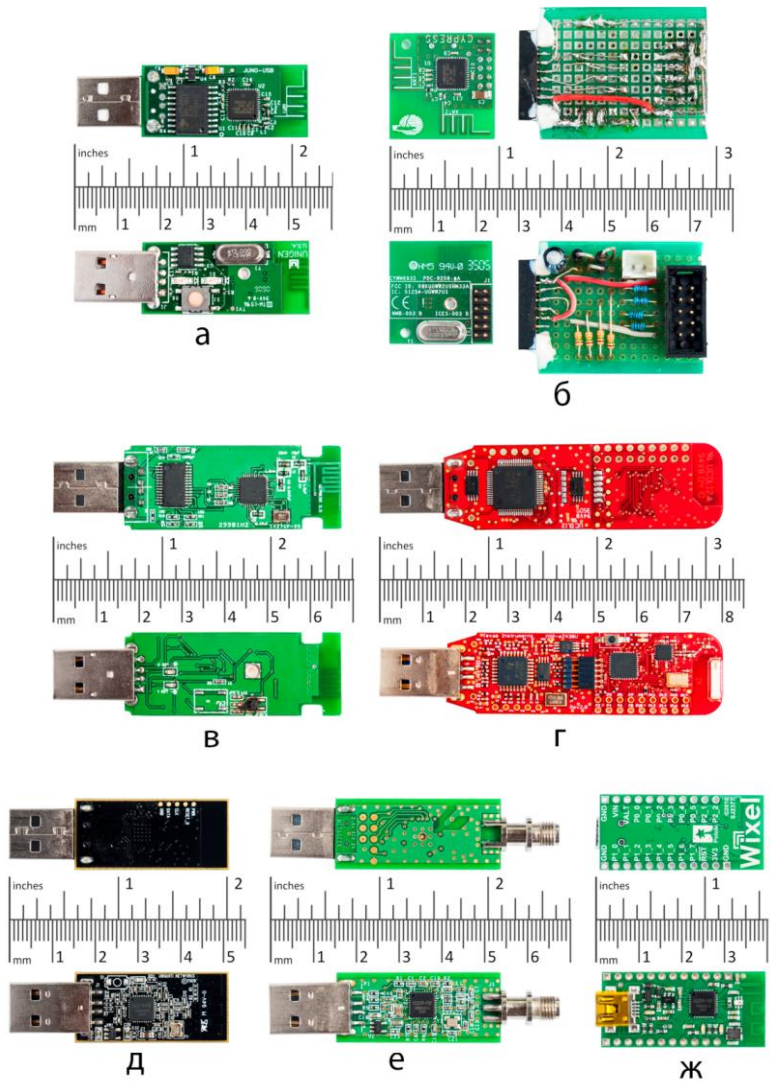

*Рис. 2. Загальний вигляд плат аналізаторів спектра:*
*а — Metageek Wi-Spy 2.4i; б — CYWUSB6935 (LPT); в — Wi-detector;*
*г — TI eZ430- RF2500; д — Ubiquiti AirView2;*
*е — Metageek Wi-Spy 2.4x і ж — Pololu Wixel*

При дослідженні ефекту затінення ПМЛ [11] комплекс засобів моніторингу доступності був виконаний на модулях Popolu Wixel, яка працює за стандартом IEEE 802.15.4 з MSK. У передавальних і прийомних модулях вбудовані планарні антени. У цьому випадку кількість точок контролю доведено до шести. Завдяки цьому вдалося кількісно оцінити область затінення, створювану ПМЛ, яка обслуговує одного з абонентів по відношенню до інших абонентам.

Останній варіант побудови комплексу для експериментальних досліджень і моніторингу параметрів електромагнітного поля, пропускної здатності, а в підсумку — доступності та цілісності інформації передбачається доповнити в кількісному відношенні (кількість датчиків) і шляхом удосконалення програмної складової прошивки датчиків для прискорення розрахунків обчислення швидкого перетворення Фур'є і Python-скриптів для аналізу даних від декількох датчиків в режимі реального часу. У перспективі це дозволить дослідити більш складні варіанти побудови систем безпроводового доступу з управлінням просторового розподілу доступності на основі застосування спрямованих антенних систем.

Крім того, досліджувалася можливість використання поляризаційного рознесення каналів або з метою двоканальної по поляризації паралельної передачі даних, або для збільшення числа абонентів шляхом їх поляризаційного поділу.

Дослідження впливу поляризаційних чинників здійснювалося шляхом виявлення зміни рівнів сигналів в залежності від розташування приймальної і передавальної антен передавача і приймача. Був побудований експериментальний канал зв'язку:

– точка безпроводового доступу (передавач) TP-Link TL-WR340G (апаратна версія 4 на мікроконтролері Atheros AR2317, 200 МГц зі стандартною прошивкою Stock 4.18.19.110701);

– зовнішній безпроводової адаптер (приймач) Linksys WUSB54G (апаратна версія 4 на мікроконтролері Ralink RT2500USB).

Обидві сторони були обладнані однаковими чвертьхвильовими вібраторами з поворотним механізмом, аналіз рівнів сигналів проводився в програмному аналізаторі спектра NetStumbler версії 0.4.0 [5–7].

Це дослідження показало, що нахил вперед приймає вібратора (в площині поширення) призводить до незначного підвищення рівня сигналу (на 2 дБмВт, можливо, за рахунок прийому хвилі, від якої потерпають від підстильної поверхні). З іншого боку, нахил в зворотну сторону дає зниження за рахунок зменшення коефіцієнта посилення.

Ортогоналізації приймають і передавальних антен може давати розподілу за рівнями до 15 дБмВт в приміщеннях з незначною кількістю елементів, що відображають і лише 4 дБмВт — з великою кількістю металевих конструкцій. Таким чином, для певної категорії споруд (виставкові зали, стадіони, криті спортивні споруди і т. ін.) Можливо поляризаційне поділ абонентів.

Для режиму МІМО виявилося, що в приміщеннях зі складною геометрією і великою кількістю перешкод і екранів використання багатопроменевої передачі дає значне поліпшення якості зв'язку за рахунок поляризаційного рознесення антен. Інформаційна швидкість може збільшитися приблизно в три рази [5–7].

## 4. СПРЯМОВАНІ ВЛАСТИВОСТІ БЕЗПРОВОДОВОГО ОБЛАДНАННЯ

Підвищення спрямованості антен, використовуваних в безпроводових системах доступу, може істотно розширити можливості і ефективність цих систем. Збільшиться дальність зв'язку, підвищиться доступність і цілісність інформації, зросте стійкість і розширяться можливості щодо застосування методів завадозахищенності, знизиться рівень взаємного впливу засобів і частка випромінювання, які не потрапляють на приймальну антену, а перетворюються на своєрідний електромагнітний сміття. Застосування спрямованого випромінювання в бік конкретного користувача підвищує його захищеність від несанкціонованого підключення інших користувачів. Якщо збільшувати спрямованість випромінювання коштів самого користувача, то зменшується ризик перехоплення переданих їм повідомлень.

Однак виникає протиріччя між необхідністю рівномірного покриття всієї обслуговуваної зони засобами базових станцій або роутерів (в припущенні, що місце розташування користувачів невідомо) і прагненням використовувати просторове розділення абонентів, яке вимагає знання його кутового розташування. Першим кроком у вирішенні цього питання для випромінювання в мережах стільникового зв'язку секторного покриття зони в азимутальній площині. Застосування більш досконалих антенних систем, наприклад, фазованих антенних решіток (ФАР) поки розглядається на теоретичному і експериментальному рівні [12–16] проте принципових перешкод на шляху їх впровадження немає. Найбільш перспективним і дає широкі можливості є застосування багатопроменевих ФАР. Питання скоріше організаційний і економічний. Основними можна вважати два варіанти: секторні ФАР і кільцеві ФАР. Останні цілком прийнятні для точок доступу систем Wi-Fi, для чого існують реальні передумови, наприклад, різновиди роутерів набором розташованих по периметру антен Linksys WRT1900AC (4 антени), D-Link DIR-895L/R AC5300 (8 антен), Asus RT-AC5300U (8 антен) і ін. Фактично, залишається один крок по модифікації їх в ФАР, збільшивши кількість випромінювачів і доповнивши систему управління взаємними фазами. Це збільшить складність і ціну обладнання, його габарити. При переході на більш високі робочі частоти ця задача полегшується.

В цілому ж впровадження різновидів ФАР в точках доступу відкриває широкі можливості по використанню адаптивних методів як з управління орієнтацією та формою діаграми спрямованості окремих каналів (аж до супроводу користувача за кутовими напрямками), так і по компенсації активних перешкод різного походження. Крім того, можливе подальше вдосконалення організації мобільного доступу, забезпечивши поділ абонентів по дальності [17, 18].

Більше технічних проблем викликає використання спрямованих антен в обладнанні користувачів через: малих розмірів гаджета, невизначеності місця знаходження користувача щодо базової станції (напрямок на роутер визначати дещо простіше); випадкової орієнтації гаджета в просторі (з ноутбуком дещо простіше); обмежень (об'єктивних і суб'єктивних) на габарити обладнання. Але і в цьому напрямку роботи ведуться. Деяке підвищення коефіцієнта посилення антени мобільного телефону можливо за рахунок її модернізації [19].

Можна запропонувати і більш радикальний, але порівняно складний варіант, - використання конформної антеною решітки (а в перспективі конформної ФАР), випромінювачі якої можна розташувати в одязі користувача, наприклад, в якійсь спеціальній куртці. На частоті 2,4 ГГц це дозволить мати в горизонтальній площині решітки з 3–4 елементів, а на частоті 5 ГГц — 5–7 (у вертикальній площині елементів може бути більше). Зрозуміло, виникають труднощі із забезпеченням фазування випромінювачів, управлінням орієнтацією максимуму характеристики спрямованості, зростає ціна і т. ін., Але для ряду ситуацій такий варіант може бути корисний (наприклад, для віддалених від базової станції туристів, альпіністів, рятувальників). В умовах безпроводового обслуговування офісних комп'ютерів особливих труднощів з розміщенням ФАР не передбачається.

Додатковим аргументів на користь застосування спрямованого випромінювання є необхідність підвищення ефективності використання випромінюваної потужності. На жаль лише незначна частина сприймається приймальною антеною. Якщо ввести коефіцієнт використання потужності як відношення прийнятої до випромінювання, то він виявиться катастрофічно малим. Використовуючи класичні співвідношення [20] Отримуємо для вільного простору $K_{исп. р} = [G(\theta,\varphi) \cdot A_{ефф}(\theta,\varphi)] / 4\pi R^2$, де $G(\theta,\varphi)$ — коефіцієнт посилення передавальної антени; $A_{ефф}(\theta,\varphi)$ — ефективна площа приймальної антени, яка прямо пропорційна коефіцієнту посилення приймальні антени; $R$ — відстань між абонентами. Тоді, наприклад, при використанні штатних антен в системі Wi-Fi з коефіцієнтами посилення порядку 2 величина $K_{исп. р}$ на дальності 10 м складе приблизно $3,8 \cdot 10^{-5}$.

З наведених формул видно, що одним із способів зменшення втрат потужності є використання більш спрямованих антен. Однак зменшити втрати на кілька порядків таким шляхом не вдасться. Виникає питання про її утилізації. Деякі приклади цього є. Наприклад, в [21] пропонуються технічні рішення, спрямовані на перетворення енергії електромагнітної хвилі в електричний струм. Це можуть бути структури, подібні до сонячних батарей на основі напівпровідникових технологій або розподілені антенні структури. Найпростіший варіант утилізації енергії ЕОМ — радіопоглинаюче покриття внутрішніх поверхонь приміщень, що перетворює енергію ЕОМ в теплову (в холодну пору року це може бути корисним).

## 5. ПЕРСПЕКТИВИ ПОДАЛЬШИХ ДОСЛІДЖЕНЬ

Автори планують продовжити дослідження прискорюють лінзових антен для підвищення спрямованості випромінювання як в сторону точки доступу, так і в бік користувача, що обґрунтовано теоретично [2, 3] і підтверджено експериментально [4–10]. Даний підхід є менш функціональний в порівнянні з ФАР, але дешевше і більш простий. Він не вимагає конструктивного зміни штатного обладнання. Однак цей спосіб дозволяє поліпшити умови доступу обмеженому числу віддалених абонентів (в силу ефекту затінення лінзи) [11]. В раді випадків це цілком прийнятно, наприклад, в офісі з не дуже щільним, але віддаленим розташуванням абонентів, що показали дослідження ефекту затінення лінз. Проведені експериментальні дослідження в [3] і [4] показали, що потужність електромагнітної хвилі в точці прийому з використанням ПМЛ збільшилася в середньому на 5–7 дБмВт, а пропускна здатність — на 4%. Дальність зв'язку збільшується в 1,8–2,2 рази при фіксованому якості, або досягається збільшення швидкості передачі при фіксованій дальності. Застосування ПМЛ в системах МІМО вимагає модифікації профілю лінзи [10]. Найбільш доцільним є установка ПМЛ близько антен користувача (мінімізується вплив на інших користувачів, індивідуально збільшується дальність). Найближчі експерименти будуть пов'язані з дослідженням впливу ПМЛ на розподіл поля у вертикальній площині і на спектральну структуру сигналу. Це можливо завдяки згаданому вище удосконалення вимірювального комплексу.

Ще один шлях збільшення дальності доступу без конструктивних змін штатного обладнання може бути заснований на екрануванні парних (особливо другій) зон Френеля [12]. Це дозволяє досягти підвищення напруженості поля в точці прийому приблизно в півтора рази. Стосовно до безпроводових систем цей спосіб підвищення доступності вимагає додаткових теоретичних і експериментальних досліджень. При невеликому числі і розосередженому за кутовими напрямками розташуванні користувачів він цілком прийнятний. Оціночні розрахунки розмірів екрану другої зони Френеля для засобів, що працюють в діапазоні 2,4 ГГц показують, що розміри таких екранів не перевищуватимуть по зовнішньому діаметру 2,5 метрів. При цьому для абонента на відстані 50 метрів кутовий конус затінення інших абонентів становитиме приблизно 4°. Наближення такого екрану зменшує його діаметр і зменшує кутовий розмір конуса затінення, що дозволяє мінімізувати ефект затінення такого екрану на сусідніх абонентів. Якщо між абонентом і точкою доступу є стіна, то на ній не складно розмістити такий екран без захаращення простору.

Експериментальна оцінка розглянутих вище методів підвищення доступності цілком можна реалізувати і планується з використанням наведених вище засобів моніторингу.

## 6. ВИСНОВКИ

Так як існуючі системи безпроводового доступу вимагаю вдосконалення, то одним з актуальних напрямків є підвищення спрямованих властивостей засобів випромінювання і прийому. Найбільш перспективним є впровадження ФАР, що не у всіх випадках може бути реалізовано в повній мірі.

Автори пропонують більш прості і менш витратні варіанти підвищення спрямованості випромінювання на основі ПМЛ, що випробувано експериментально, а також використання зонування простору на шляху ЕОМ. Обидва варіанти не вимагають конструктивних змін штатних засобів систем Wi-Fi.

Для експериментальних досліджень запропонованих варіантів розроблений вимірювальний комплекс, що дозволяє оцінювати просторовий розподіл рівнів випромінювання, їх спектральну структуру і пропускну здатність каналів.

## СПИСОК ВИКОРИСТАНИХ ДЖЕРЕЛ

**Volodymyr M. Astapenya**
PhD, associate professor
Borys Grinchenko Kyiv University, Kyiv, Ukraine
OrcID: 0000-0003-0124-216X
v.astapenia@kubg.edu.ua

**Volodymyr Yu. Sokolov**
MSc, senior lecturer
Borys Grinchenko Kyiv University, Kyiv, Ukraine
ORCID: 0000-0002-9349-7946
v.sokolov@kubg.edu.ua

**Mahyar TajDini**
MSc, junior researcher
Borys Grinchenko Kyiv University, Kyiv, Ukraine
OrcID: 0000-0001-8875-3362
m.tajdini@kubg.edu.ua

# RESULTS AND TOOLS FOR EVALUATING THE EFFECTIVENESS OF FOCUSING SYSTEMS TO IMPROVE ACCESSIBILITY IN WIRELESS NETWORKS

**Abstract.** The widespread use of wireless technologies leads to an ever-increasing number of users and permanently functioning devices. However, the growth of the number of wireless users in a limited space and a limited frequency range leads to an increase in their mutual influence, which ultimately affects the throughput of wireless channels and even the performance of the system as a whole. The article presents the statistics and tendencies of the distribution of wireless networks of the IEEE 802.11 standard systems, as well as analyzes the main problems that arise during the expansion of their use. Substantiation and choice of ways to overcome these difficulties largely depends on the objective control of radiation parameters of access points and subscriber funds in a particular environment. The review of the state control facilities provided by the developers of the equipment is presented, and author's variants of experimental measuring complexes are offered, allowing to control signal and information parameters of Wi-Fi systems. The experimental results obtained with the use of the indicated means, obtained using the accelerating metal-plate lens as an additional autonomous element for focusing the field, including for MIMO systems, the effect of the accelerating metal-plate lens on the spatial distribution of the field, on the spectral structure of the signal are presented. In addition, polarization effects were investigated. Possible ways to further increase the availability, integrity of information and energy efficiency of wireless access systems are discussed. The authors propose simpler and less costly options for increasing the direction of radiation on the basis of an accelerating metal-plate lens, experimentally tested, as well as the use of zone zoning on the path of the computer.

**Keywords:** wireless network; metal-plate lens; accelerating lens; lens antenna; access point; polarization; MIMO.